\documentclass[aps,pre,longbibliography,notitlepage,floatfix]{revtex4-1}

\usepackage{bm,amsmath,amssymb,graphicx}
\usepackage[abs]{overpic}
\newcommand{\uvc}[1]{\bm{\mathrm{\hat #1}}} 

\newcommand{\bX}{{\bf X}}
\newcommand{\bg}{{\bf g}}

\begin{document}

\title{Slack Dynamics on an Unfurling String}
\author{J. A. Hanna}
\email{hanna@physics.umass.edu}
\author{C. D. Santangelo}
\email{csantang@physics.umass.edu}
\affiliation{Department of Physics, University of Massachusetts, Amherst, MA 01003}

\date{\today}

\begin{abstract}
An arch will grow on a rapidly deployed thin string in contact with a rigid plane.  We present a qualitative model for the growing structure involving the amplification, rectification, and advection of slack in the presence of a steady stress field, validate our assumptions with numerical experiments, and pose new questions about the spatially developing motions of thin objects.
\end{abstract}

\maketitle

\begin{figure}[here]
\includegraphics[width=4.3in]{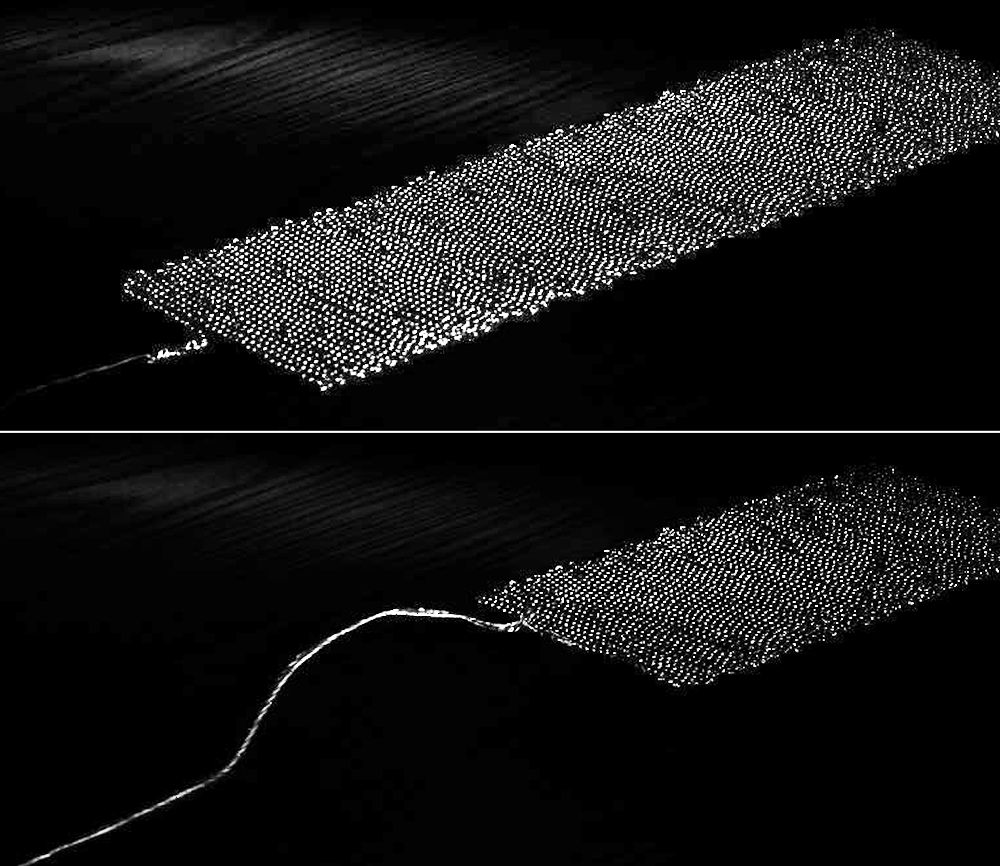}
\caption{Still images from a video \cite{HannaKing11} showing configurations of a (real) chain compactly arranged on a table before (top) and during (bottom) a straightening process.  A vertically-oriented arch is apparent in the lower image.  Chain links are $\sim$2 mm, the amplitude of the initial condition is $\sim$5 cm, the imposed velocity of the pulled end is $\sim$8 m/s.}
\label{arch}
\end{figure}

The deployment of a quasi-one-dimensional object from a roll or pile is a canonical event, important for textile manufacture \cite{Padfield58,KothariLeaf79both,Fraser92}, tethered satellite control \cite{BeletskyLevin93,MankalaAgrawal05,Krupa06}, intercontinental telecommunications \cite{Thomson1857-1,*Thomson1857-2,Airy1858,Zajac57}, aerial refueling, boating, and gardening.  Despite this ubiquity, we were recently surprised by an observation of these dynamics, as embodied by a rapidly straightened chain on a tabletop \cite{HannaKing11}.  The surprise in question is displayed in Figure \ref{arch}.  One free end of an orderly monolayer of chain is rapidly pulled along a table, leading to the formation of a slowly growing, wobbling, noisy arch near the pick-up point.  The phenomenon is not unique to chains, and may be observed in strings, ropes \cite{MythB}, and similar objects; we will use the words ``chain'' and ``string'' interchangeably in what follows.  We will suggest a mechanism for this robust behavior by considering the simplest analytical and numerical systems resembling that of Figure \ref{arch}: an inextensible string, and a chain of beads and springs.

Given a time-dependent curve $\bX(s,t)$ parametrized by arc length $s$, the wave equation
\begin{equation}\label{vectorwave}
	\mu\partial^2_t \bX= \partial_s\left(\sigma\partial_s \bX\right)
\end{equation}
describes a balance of inertia and line tension in a string of uniform mass density $\mu$.  The stress $\sigma(s,t)$, morally equivalent to the pressure in a one-dimensional fluid, is a multiplier field enforcing the metrical constraint $\partial_s\bX\cdot\partial_s\bX=1$ \cite{EdwardsGoodyear72,Hinch76,Hinch94,Reeken77,Healey90,GoldsteinLanger95,Thess99,ShelleyUeda00,Belmonte01,SchagerlBerger02,Preston11-2}.  This equation is, in principle, derivable as the zero-radius limit of an elastic rod, or the continuum limit of a chain of rigid links.  Taking an arc length derivative of \eqref{vectorwave} and projecting along the unit tangent $\partial_s\bX$ yields the auxiliary equation
\begin{equation}\label{constraint}
	\partial^2_s\sigma - \sigma\kappa^2 = -\mu\partial_t\partial_s\bX\cdot\partial_t\partial_s\bX \, ,
\end{equation}  
the solution of which serves to impose the constraint \cite{EdwardsGoodyear72,Hinch76,GoldsteinLanger95,Thess99,ShelleyUeda00,Belmonte01,SchagerlBerger02,Preston11-2}.  The ``screening potential'' of this Poisson-like equation is the squared curvature $\kappa^2 = \partial^2_s\bX\cdot\partial^2_s\bX$, while the ``source'' term on the right hand side has been rewritten in terms of the magnitude of the rate of change of the unit tangent using $\partial_t\partial_s\bX\cdot\partial_s\bX=0$.  From this form, it appears that the stresses induced by inertial motions of the string will be tensile unless one imposes compression through boundary or initial conditions.  Positivity of the stress is proven for one fixed and one free end \cite{Preston11-2}, and the proof may be extended to other conditions \cite{Preston12pc}.  Hence, we will not generally be concerned with unstable evanescent dynamics that might arise from imaginary wave speeds in \eqref{vectorwave}.

Relevant to our problem are a class of solutions to \eqref{vectorwave} and \eqref{constraint} for which the curve motion is a pure tangential velocity $T$, so that $\partial_t\bX = T\partial_s\bX$.  Here, $\bX$ is arbitrary and $T$ and $\sigma = \mu T^2$ are uniform along the curve \cite{Routh55,HealeyPapadopoulos90}.  The shape $\bX$ itself is simply one of two waves, namely that which travels at speed $-T$ with respect to the Lagrangian frame of \eqref{vectorwave} moving at speed $T$ along the curve.  It is thus stationary in the laboratory frame; this and the linear dispersion inherent when $\sigma$ is uniform are used to great effect in certain kinetic sculptures \cite{Aitken1876,LariatChain,StringLauncher}.

How important is gravity?  The addition of a uniform term $\mu\bg$ to the right hand side of \eqref{vectorwave} breaks the degeneracy in $\bX$ and may generate compressive stresses.  The remaining stationary-shape solutions of this augmented equation are the catenaries \cite{Airy1858, Routh55, PerkinsMote87,SchagerlBerger02} with the tangential velocity $T$ determining the tension, not the shape.  Given that compression ($\sigma < 0$) corresponds to catastrophic failure, we can naively assume inertial stabilization of a symmetric catenary arch $\kappa = \frac{a}{a^2s^2 + 1}$, $a$ a negative constant, bearing a tension $\sigma = \mu\left(T^2 + \| \bg \| \frac{\sqrt{a^2s^2+1}}{a}\right)$, as long as the length parameters of the shape fall within a circle whose radius is the natural length scale defined by the velocity and gravity: $\left(\frac{1}{a}\right)^2 + s^2 \le \left(\frac{T^2}{\| \bg \|}\right)^2$.  At this point, it is worth recalling the scales involved in Figure \ref{arch}.  The terminal ``free stream'' tangential velocity $T \sim$ 8 m/s, $\bg$ is Earth gravity ($\sim$ 10 m/s$^2$), and $\kappa \ge\, \sim$ 10 /m for the duration of the experiment.  Hence, the ratio of inertial to gravitational accelerations, $\frac{T^2 \kappa}{\| \bg \|}$, is at least on the order of fifty to a hundred.  So gravity is neglectable in this system, due to the large curvature that spontaneously emerges early in the process.  Gravity will only become relevant given sufficient growth time, or a stiffer curve with a bending or twisting length scale that could set a lower bound on the initial emergent curvature.  Adding a rigid motion with comparable velocity to the analysis--- a wobbling perpendicular to the catenary's plane, for example--- does not change the order of magnitude of the relevant terms.  

\begin{figure}[here]
\includegraphics[width=4.3in]{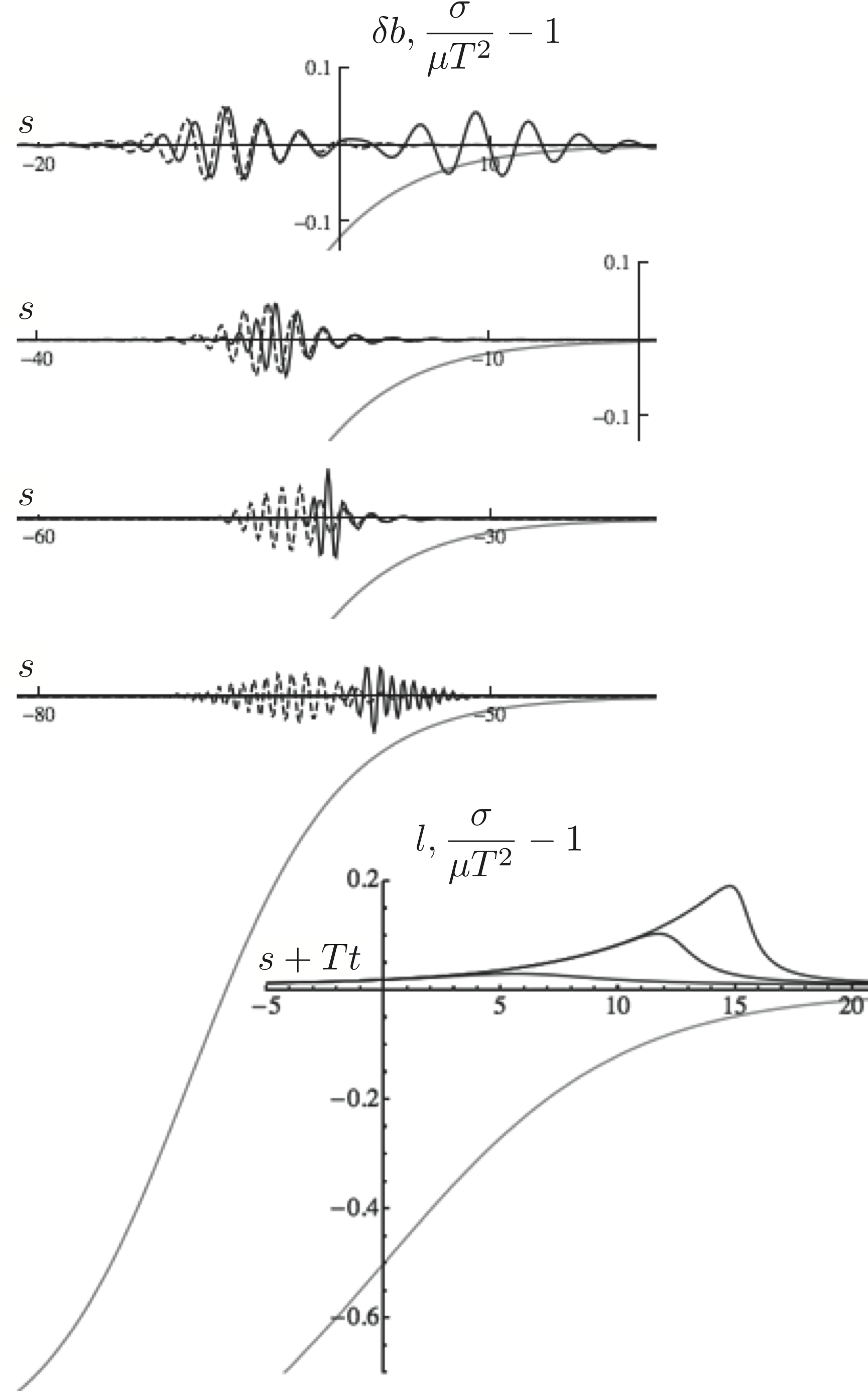}
\caption{The grey curves on all five plots are the same normalized stress, shifted by unity for visual convenience: $\frac{\sigma}{\mu T^2}-1$, where $\sigma=\frac{\mu T^2}{2}\left[1+\tanh\left(\frac{\eta}{10}\right)\right]$ and $\eta \equiv s+Tt$.  TOP FOUR PLOTS: The evolution of a wave packet $\delta b(s,t)$ under \eqref{scalarwave} (full black curves) and the integrated form of \eqref{slow} (dashed black curves), for $\mu=T=1$, at times $t=10, 30, 50, 70$, as viewed in a frame Eulerian on the string, with the downstream direction to the right and the horizontal axes representing $s$.  The initial conditions at $\eta=s$ are $\delta b = 2b_0$ for \eqref{scalarwave} and $\delta b = b_0$ for \eqref{slow}, where $b_0 \equiv \frac{1}{20}\mathrm{sech}\left(\frac{3\eta}{10}\right)\cos\left(2\eta\right)$.  LOWER RIGHT INSET: The evolution of the slack variable $l \equiv \sqrt{\left(\partial_s\delta b\right)^2}$ under \eqref{slack} (full black curves) for $\mu=T=1$, at times $t=50, 150, 250$, as viewed in a frame Eulerian on the string, with the downstream direction to the right and the horizontal axis representing $\eta \equiv s+Tt$.  The initial condition at $\eta=s$ is uniform: $l = \frac{1}{100} .$ }
\label{twotime}
\end{figure}

Returning to our original equations \eqref{vectorwave} and \eqref{constraint}, let's consider an initially planar curve $\bX$ with a small perturbation $\delta b\uvc{b}$ oriented perpendicularly to the plane containing $\bX$. To first order in the perturbation, we have the decoupled equations
\begin{eqnarray}
	\mu\partial^2_t\delta b &=& \partial_s\left(\sigma\partial_s\delta b\right) \label{scalarwave} \, ,\\
	\partial^2_s\delta\sigma - \delta\sigma\kappa^2 &=& 0 \, ,	
\end{eqnarray}
which imply that $\delta\sigma = 0$ and the height function $\delta b$ moves through a stress background determined entirely by the planar dynamics.  What should this stress profile look like for an unfurling planar string?  Modulo the periodic component of the motion induced by the initial chain layout, material particles take heteroclinic trajectories between two arbitrary stationary shapes with tangential velocities of zero in the pile and the imposed pulling velocity $T$ downstream.  These shapes correspond to stresses of zero and $\mu T^2$, respectively.  Although one might expect velocity and stress discontinuities right at the moving front, we do observe a smooth transition region for the velocity in the experiment \cite{HannaKing11}, which presumably corresponds to a smooth transition in stress.  A nonperiodic, or temporally averaged, part of the dynamics appears to be nearly steady; the arch grows slowly and remains within a nearly constant distance from the front.  Additionally, equations \eqref{vectorwave} and \eqref{constraint} indicate that time-dependent curves correspond to nonuniform, but not necessarily time-dependent, stress profiles; large periodic deformations of the string do not imply similar features in the stress.  Thus, we introduce the traveling wave variable $\eta \equiv s+Tt$ and provisionally presume a steady stress $\sigma(\eta)$ in the corresponding frame.  In this frame, Eulerian with respect to the curve but not the laboratory, we may approximate the slowly evolving dynamics \eqref{scalarwave} with the equation
\begin{equation}\label{slow}
	\partial_\eta\left( \partial_t \delta b - \frac{\sigma(\eta) - \mu T^2}{2\mu T}\partial_\eta \delta b \right) =0 \, .
\end{equation}
This is a short-time description of $\delta b(\eta, t)$ that presumes smallness of second order time derivatives.  It admits general solutions, for which the characteristics converge at infinite time, of the form $\delta b = f\left(t + \int^\eta d\eta' \frac{2\mu T}{\sigma(\eta')-\mu T^2}\right)$, $f$ an arbitrary function.  Figure \ref{twotime} compares the evolution of a wave packet under \eqref{scalarwave} and the integrated form of \eqref{slow} for a tanh-shaped $\sigma(\eta)$ that interpolates between zero and the free stream stress $\mu T^2$.  Agreement is good for early times, when the packet is squeezed into a smaller footprint behind the free stream.  The disturbance eventually reaches a nearly stationary configuration \emph{via} a complicated modulated motion, not captured by the first order transport equation \eqref{slow}.  Note that squeezing the packet does not increase its amplitude.

These transient, geometrically linearized dynamics are best expressed in a new variable, the ``slack'' $l \equiv \sqrt{\left(\partial_s\delta b\right)^2}$ taken up into the height function.  In terms of $l$, \eqref{slow} takes the form
\begin{equation}\label{slack}
	\partial_t l = \partial_\eta \left(  \frac{\sigma(\eta) - \mu T^2}{2\mu T} l \right) \, .
\end{equation}
Given a monotonically increasing $\sigma(\eta) < \mu T^2$, the right hand side of \eqref{slack} is composed of an amplification and an advection term.  These are the ingredients for a convectively unstable system \cite{Deissler85,DeisslerFarmer92,HuerreMonkewitz90,Chomaz05}.  However, in contrast to the open flow systems usually considered in this context, both the amplification and advection coefficients in \eqref{slack} drop to zero as the material particle moves downstream through a stress field that smoothly approaches its free stream value.  The slack pools behind the free stream, as shown in the lower inset of Figure \ref{twotime}.  This should correspond to a location near the upstream base of the arch.  

In the experiment, this transient, localized slope amplifier was coupled to a height rectifier, namely a table.  Is this sufficient for the formation of an arch?  To explore this question, we have implemented a simple Verlet integration \cite{FrenkelSmit96} of massive beads and stiff springs.  Thus, while introducing extensibility, we exclude bending and twist elasticity, air drag, frictional interactions with the table, and any other messy and experimentally unavoidable physics.  The 5000 beads begin sinusoidally arranged in a plane defined by the potential minimum created by a constant ``gravity'' perpendicular to a hard plane ``table'', an exponential with argument 100 at a penetration depth of one spring rest length.  A velocity in the direction of the sine's argument is imposed on one end bead.  A small quantity of uniform noise is added to the vertical force density on each bead at all times, and there is also a damping coefficient which falls exponentially towards zero from an initial value of a few percent.

\begin{figure}[here]
\includegraphics[width=5in]{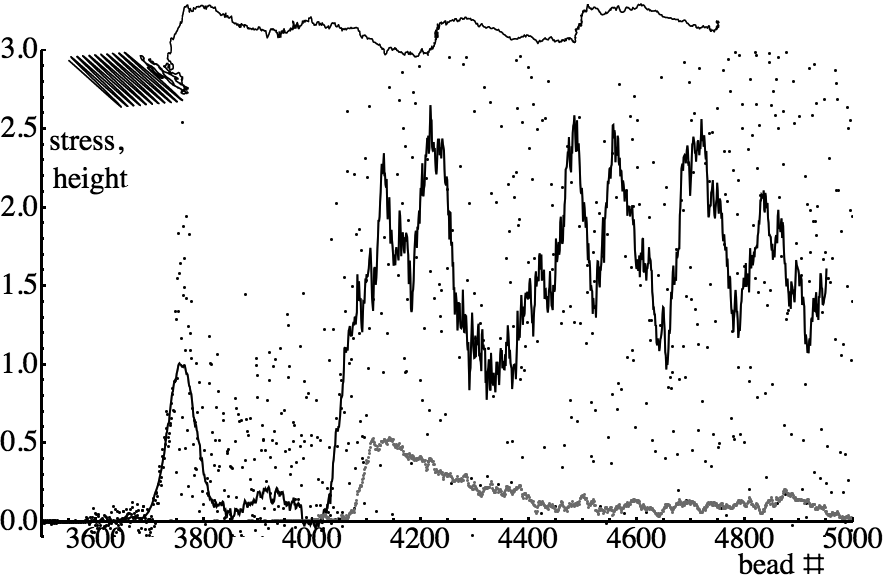}
\caption{Numerical results.  Height (grey curve, normalized by the initial layout amplitude), stress (black dots), and averaged stress (black curve, moving average over 50 beads, normalized by the theoretical free stream stress $\mu T^2 = 1$) after 150 ms for a configuration with initial layout amplitude and frequency of $\sim$6 cm and $\sim$0.6 cycles/cm.  The velocity boundary condition is imposed on bead \#5000.  A portion of the corresponding three-dimensional shape is shown in an oblique view in the upper left inset.  Spring rest lengths are 2 mm, and the strain corresponding to a stress of 1 is 2.5\%.  The raw stress data extend from approximately -9 to +11 in the noisy plateau region.  Noise from a uniform distribution of half-width $10^{-3}$ is added to vertical force densities; this may be compared to $\mu\| \bg \| = 1$ and spring force densities on the order of $10^7$.}
\label{mimic}
\end{figure}

The snapshot of shape, stress, and height in Figure \ref{mimic} is the result of a simulation with initial layout amplitude and frequency of $\sim$6 cm and $\sim$0.6 cycles/cm, parameters similar to Figure \ref{arch}.  We have set $\mu = \| \bg \| = T = 1$, and the half-width of the uniform noise distribution is $10^{-3}$.  The extensions of the springs are used as a proxy for stress, and these data are spatially averaged.  An initial attenuating pulse travels down the chain, leaving a nearly quiescent wake, followed by a rapid transition to the free-stream stress $\mu T^2$.  This very noisy stress plateau also carries oscillations due to the undamped lateral motions of the chain arising from its initial configuration.  An arch does indeed form, on the downstream side of the transition.  The effect of gravity in the simulation is rather minimal, affecting only the qualitative shape of the arch.  However, in simulations without the table and gravity potentials, the slack isn't rectified, and no arch forms.  The dynamics are quite robust with respect to changes in the spring constant.  The simulations shown in this letter involve strains of about 2.5\% in the plateau, but the important features are quite similar for strains an order of magnitude larger or smaller.  We note in passing that the damping term in the dynamics causes effects comparable to those of a noise source.  The qualitative behavior shown here persists as long as the noise is above a threshold value, which is on the order of $10^{-8}$-$10^{-9}$ in the absence of damping.

\begin{figure}[here]
\includegraphics[width=5.5in]{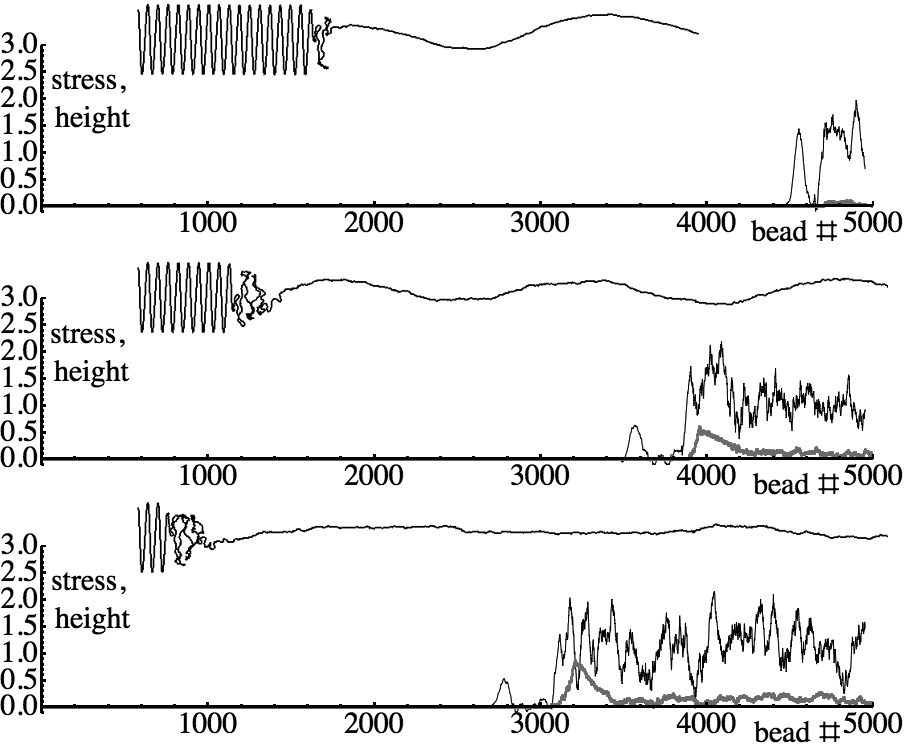}
\caption{Numerical results.  Height (grey curves, normalized by the initial layout amplitude) and averaged stress (black curves, moving average over 50 beads, normalized by the theoretical free stream stress $\mu T^2 = 1$) after 55, 175, and 295 ms for a configuration with initial layout amplitude and frequency of $\sim$6 cm and $\sim$0.6 cycles/cm.  Top views of a portion of the corresponding shapes are shown in the upper insets.  Other details as for Figure \ref{mimic}.}
\label{pullhi}
\end{figure}

\begin{figure}[here]
\includegraphics[width=5.5in]{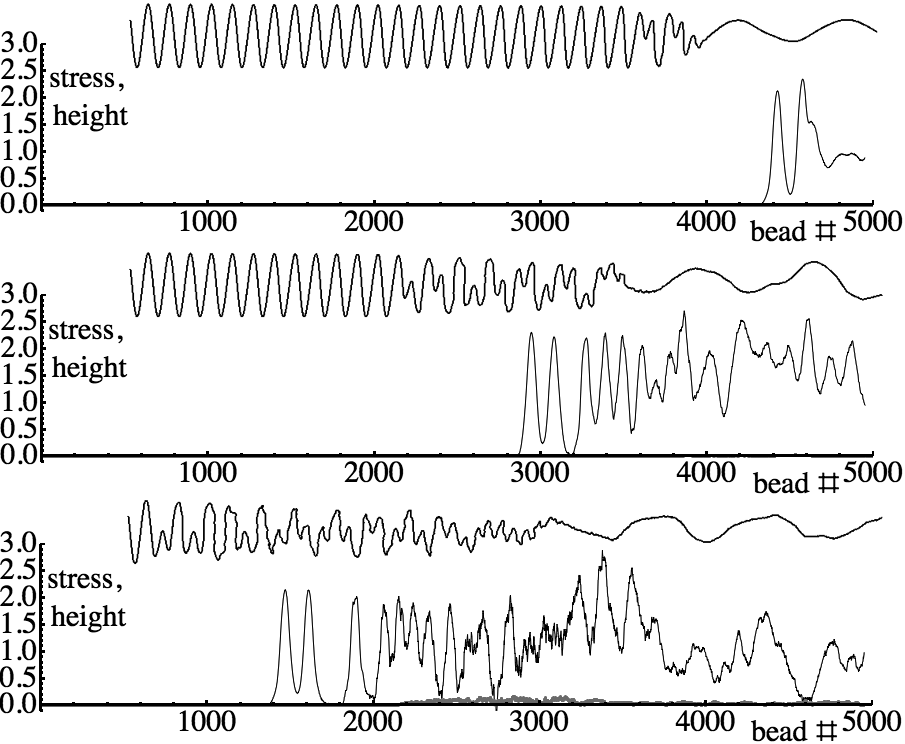}
\caption{Numerical results.  Height (grey curves, normalized by the initial layout amplitude) and averaged stress (black curves, moving average over 50 beads, normalized by the theoretical free stream stress $\mu T^2 = 1$) after 55, 175, and 295 ms for a configuration with initial layout amplitude and frequency of $\sim$6 cm and $\sim$0.1 cycles/cm.  Top views of a portion of the corresponding shapes are shown in the upper insets.  Other details as for Figure \ref{mimic}.}
\label{pulllo}
\end{figure}

We return to our assumption of a steady stress distribution in the traveling wave frame.  Figures \ref{pullhi} and \ref{pulllo} show snapshots of two chains at three identical times.  The chain of Figure \ref{pullhi} is that of Figure \ref{mimic}, while that of Figure \ref{pulllo} corresponds to an initial arrangement with similar amplitude but lower frequency, $\sim$6 cm and $\sim$0.1 cycles/cm, and the same level of added noise.  The high-frequency chain appears to sustain a steady, or perhaps very slowly growing, transition region between low and high velocity, and is highly receptive to noise.  The transition region moves approximately 6-7 beads/ms upstream, not much faster than the traveling wave coordinate (5 beads/ms).  The transition region on the low-frequency chain steadily widens as the chain unfurls, its leading edge moving significantly faster upstream (9-11 beads/ms), and out-of-plane growth is slower to develop and less pronounced.  Over the time interval shown, this region retains much qualitative information about its initial shape even as the material within it has achieved highly tangential velocities.  The stress on this chain appears to reflect this ordered arrangement, but the nature of the pulse train in Figure \ref{pulllo} remains unknown.  The validity of our steady stress assumption apparently depends on the initial conditions, specifically the curvature, or screening, of the string.  It is also worth noting that, for both of the initial layouts shown, there are regions of chain in which the averaged stress becomes compressive for brief periods.  These appear to behave as sources of slack for the system to amplify.

Our preliminary study has raised several questions.  Perhaps most interesting is the issue of how to analyze the planar dynamics we see in the insets of Figures \ref{pullhi} and \ref{pulllo}, and the associated stress generation, in terms of the time-dependent screening and source terms of equation \eqref{constraint}.  There is some sort of signal being transmitted along our string, conveying information about the velocity boundary condition.  The process of transmission is highly dependent on the evolving shape of the curve, and thus more complicated than transverse impulsive loading of a straight string \cite{BeattyHaddow85}.  Interesting physics arises from gradients in the stress, which are absent in the Klein-Gordon-type equations describing the perturbative dynamics around stationary states of elastic rods \cite{GorielyTabor97-2}, though we note that a uniform but time-varying tension is known to cause instabilities of planar motions of fixed-end strings \cite{Gough84}.  What slowly evolving stress distributions can be sustained by moving elastic curves? And which spatially developing motions will inevitably bear structures like our arch?  Are there analogous phenomena in elastic sheets, nets, fluid membranes, or the Navier-Stokes equations in certain geometries?  Incompressibility of the system appears merely an analytical convenience, not a strict requirement.  Do short-lived zones of compressive stress only appear in compressible systems, and are they important?  Finally, we have disregarded the periodic aspects of the curve's motion, and still lack an explicit description of the role of the rectifying potential.  Is it possible to describe the arch growth as a secular result of some forcing by the table, perhaps on the highly curved turning points, or is the resemblance to a resonance superficial?

\clearpage

\section*{Acknowledgments}

We thank H.\ King and N.\ Menon for help, comments, and encouragement, and J. Machta for a discussion.  JAH thanks B.\ Mbanga for advice on simulations, R. Schroll for a two-timing tutorial, P. Kevrekidis for a discussion, and B.\ Eckhardt for a suggestion about frames.  Funding came from National Science Foundation grant DMR 0846582.

\bibliographystyle{unsrt}


\end{document}